\documentclass[twocolumn,superscriptaddress,showpacs,aps,prb]{revtex4}
\usepackage{mathtext}
\usepackage{graphics}
\usepackage[dvips]{epsfig}
\usepackage{amsmath}
\usepackage{amsmath}
\def\Xint#1{\mathchoice
{\XXint\displaystyle\textstyle{#1}}%
{\XXint\textstyle\scriptstyle{#1}}%
{\XXint\scriptstyle\scriptscriptstyle{#1}}%
{\XXint\scriptscriptstyle\scriptscriptstyle{#1}}%
\!\int}
\def\XXint#1#2#3{{\setbox0=\hbox{$#1{#2#3}{\int}$}
\vcenter{\hbox{$#2#3$}}\kern-.55\wd0}}

\def\dashint{\Xint-}
\newcounter{fig}

\begin{document}
\title{ Influence of ferromagnetic ordering on  Raman scattering in CoS$_2$}
\author{L.A. Falkovsky}
\affiliation{ Landau Institute for Theoretical Physics, Moscow
119334\\Verechagin Institute of the High Pressure
Physics, Troitsk 142190}
\date{\today}

 \begin{abstract}
 
The effects of phonon anharmonicity, phonon-magnon and electron-phonon interactions on the temperature dependence of  Raman optical phonon modes are theoretically investigated. Besides of the Klemens result for the phonon width due to  anharmonicity, the corresponding lineshift is derived.
We argue that the phonon  decay into two magnons has very low intensity in ferromagnets with  low Curie temperatures. Therefore, the  electron interband transitions accompanied with the ferromagnetic ordering are included in considerations to get a good quantitative agreement with  experiments.

% 65.80.+n,71.70.Di, 71.18.+y}
\end{abstract}

\maketitle

\section{Introduction}

Thermal broadening of  phonon lines in the Raman scattering is usually described in terms of three-phonon anharmonicity, i.e. by the decay of an optical phonon with a frequency $\omega $ in two phonons.

The simplest case when the final state has two acoustic phonon from one branch was theoretically studied  by Klemens \cite{Kl},
who obtained the temperature dependence of the Raman linewidth. The corresponding lineshift was  considered in Refs. \cite{BWH, MC}.

This theory  was compared in works \cite{BWH,MC,DBM}  with experimental data for Si, Ge, C, $\alpha-$Sn.
 A model was also considered with the phonons in the final state from different branches. It was found that  anharmonic interactions of the forth order should  also be included 
in order to describe  precisely  the Raman line behavior at high temperatures $T> 300$ K. 

The interaction of phonons with magnons in antiferromagnets was discussed also in the analysis of the thermal conductivity \cite{MOZ},  the spin Seebeck effect \cite{UTH,JYM},  high-temperature superconductivity \cite{NMH}, and optical spectra \cite{KKP}.

 The magnon-phonon interaction   results in the magnon damping  \cite{Wo}, however, no effect for phonons was  shown.   Damping of the optical phonons was found \cite{MPB} to become large  in the rear-earth Gd and Tb below the Curie temperature achieving a value of 15 cm$^{-1}$, which is much greater than the   three-phonon interaction effect.

Temperature variation of the electronic structure of  half-metallic CoS$_2$ was investigated by means of reflectivity measurements \cite{YMM}. Recently \cite{La}, the Raman scattering in CoS$_2$ was studied at temperatures nearby the ferromagnetic transition at $T_c=122$ K. The shift and width of the $\omega=400 $ cm$^{-1}$ line are observed as functions of  temperature. It is seen that an additional mechanism of the lineshift   is incorporated around the Curie temperature.  

In this theoretical paper, we consider the different interactions of the optical phonons in order to explain their temperature dependences observed. First, we obtain both the width and shift  of the Raman line due to anharmonic interactions of the third order. Then, the interaction of  phonons with magnons below the Curie temperature is considered. At last,
the effect of the ferromagnetic ordering on the phonon-electron interaction is studied.

\section{Shift and width of the optical phonon due to anharmonic interactions of the third order}

\begin{figure}[h]
\resizebox{.2\textwidth}{!}
%\epsfxsize=40mm
 %\epsfysize=40mm
%\centerline{\epsfbox{diag.eps}}
{\includegraphics{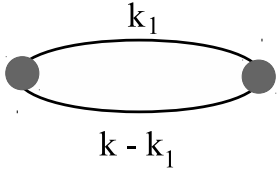}}%pw1.eps}}%xy7.eps
\caption{(Color online) Phonon self-energy; two lines of the loop represent two phonons in the case of three-phonon interactions or two magnons for the phonon-magnon interactions. }
\label{diag}
\end{figure}

Here we calculate the  Raman phonon self-energy due to three-phonon anharmonicity represented by the loop in  Fig. \ref{diag}. Two lines of the loop correspond with two phonons in the final state, and the shaded circle  shows   the interaction  vertex of these two phonons with the initial optical phonon. At given frequency $\omega$ and momentum ${\bf k}$ of the optical phonon, we  have to summarize  over the momenta ${\bf k}_1$ and ${\bf k}_2$ of  phonons in the final state and to perform the Matzubara summation over the frequencies $\omega_n=2\pi nT$. One summation over the momentum of the final phonon can be done using the conservation low $\bf{k}_1+\bf{k}_2=\bf{k}$. 

Then we meet the sum
\begin{equation}
S_{ph-ph}(\omega, {\bf k})=-T\sum_{n, {\bf k_1}}\frac{2\omega_{\bf{ k}_1}}{\omega_{{\bf k}_1}^2+\omega_n^2}\,       \frac{2\omega_{\bf{k-k}_1}}{\omega_{\bf{k-k}_1}^2+(\omega-\omega_n)^2}\,,
\label{sum}\end{equation}
where each of the two factors is the Green function of the phonon in the final state
\[D_{ph}({\bf k}_1, \omega_n)=\frac{-2\omega_{\bf{ k}_1}}{\omega_{{\bf k}_1}^2+\omega_n^2}\,.\]
We perform the summation over  $n$ considering the integral over the large circle in the complex $z$-plane   of the integrand 
$ f(i\omega_n=z)n_B(z)/2\pi i$, where $f(i\omega_n)$ is the function under the sum-sign in Eq. (\ref{sum}) and $n_B(z)=[\exp{(z/T)}-1]^{-1}$.  If the circle radius goes to the infinity, then the integral  tends obviously to zero. It means that the sum of  all the residues inside  the circle  equals  zero. The residues of the function $n_B(z)$  at the poles give the sum in Eq. (\ref{sum}), which is thus equal to the sum with the opposite sign  of  four residues at the poles  $\pm\omega_{{\bf k}_1}$ and $\pm\omega_{\bf{k-k}_1}$. For definiteness, we assume that the frequencies $\omega_{{\bf k}_1}$ and $\omega_{\bf{k-k}_1}$ are positive. At $z=\omega_{{\bf k}_1}$, the function  $n_B(z)$ coincides  with the number of phonons $N(\omega_{{\bf k}_1})$ and  it gives  $-[N(\omega_{{\bf k}_1})+1]$ at $z=-\omega_{{\bf k}_1}$.

Finally, the analytical continuation on the real frequency $\omega$ is performed by the substitution $\omega\rightarrow i\omega-\delta$ with the infinitesimal positive $\delta$. The following  four terms appear: 
\begin{equation}
\begin{array}{c}
\displaystyle{\frac{2\omega_{\bf{k-k}_1}N(\omega_{\bf{ k}_1})}{\omega_{\bf{k-k}_1}^2-(\omega_{\bf{ k}_1}+\omega+i\delta)^2}+
\frac{2\omega_{\bf{k-k}_1}[N(\omega_{\bf{ k}_1})+1]}{\omega_{\bf{k-k}_1}^2-(\omega_{\bf{ k}_1}-\omega-i\delta)^2}}\\
\displaystyle{+\frac{2\omega_{\bf{ k}_1}N(\omega_{\bf{k-k}_1})}{\omega_{\bf{ k}_1}^2-(\omega_{\bf{k-k}_1}-\omega-i\delta)^2}+
\frac{2\omega_{\bf{ k}_1}[N(\omega_{\bf{k-k}_1})+1]}{\omega_{\bf{ k}_1}^2-(\omega_{\bf{k-k}_1}+\omega+i\delta)^2}}\,.
\end {array}
\label{res}\end{equation}

We are interested in the Raman phonon with ${\bf k}=0$, and the phonon dispersion satisfies  the condition  $\omega({\bf k}_1)=\omega(-{\bf k}_1)$.
Therefore, we can combine the terms in Eq.  (\ref{res}) to get
\begin{equation} \frac{4\omega_{\bf{ k}_1}[1+2N(\omega_{\bf{k}_1})]}{4\omega_{\bf{ k}_1}^2-(\omega+i\delta)^2}\,.
 \label{resu}\end{equation}

 To obtain  the phonon self-energy, the Eq. (\ref{resu}) should be integrated over ${{\bf k}_1}$
 with the  three-phonon anharmonic vertex  squared. This vertex  appears  \cite{LP} while the three-phonon Hamiltonian is expressed in terms 
of the phonon operators giving a factor $ (\omega\omega_{{\bf k}_1}^2)^{-1/2}$. Each acoustic phonon with the momentum ${\bf k}_1$ contributes  an additional   multiplier $k_1$. Therefore,
the three-phonon vertex squared has a form $w_{ph-ph}^2=g_{ph-ph}^2k_1^2(\omega\omega^2_{\bf{ k}_1})^{-1}$, where the function $g_{ph-ph}$ is independent of the momentum $k_1$.

 The phonon self-energy writes 
 \begin{equation}
\begin{array}{c}
\Sigma_{ph-ph}(\omega,T)=\displaystyle{\frac{1}{2\pi^3\omega}\int\frac{g^2_{ph-ph}k_1^2[1+2N(\omega_{\bf{ k}_1})]d^3{\bf k}_1 }{\omega_{\bf{ k}_1}[4\omega_{\bf{ k}_1}^2-(\omega+i\delta)^2]}}\end {array}
\label{res31}\end{equation}
 with  real and imaginary parts.  

The imaginary part of Eq. (\ref{res31}) integrated over $k_1$ or over $\omega_{{\bf k}_1}=k_1/s$, where $s$ is the sound velocity, yields the Klemens formula
\begin{equation}
\Gamma_{ph-ph}(\omega,T)=\Gamma_{ph-ph}(\omega,0)[1+2N(\omega/2)]
\label{kl}\end{equation}
 with the linewidth at  zero temperature
 \begin{equation}
\Gamma_{ph-ph}(\omega,0)=\frac{\bar{g^2}_{ph-ph}\omega}{2^4\pi s^5}\,,
\label{res2}\end{equation}
where   the averaging with respect to angles is denoted by the overline.
% \begin{equation}
%\Gamma(\omega,0)=\frac{1}{8\pi^2\omega\omega_D^3}\int g^2_{ph-ph}d\Omega\sim\frac{\omega_D^2}{2\pi \omega}\left(\frac{m}{M}\right)^{1/2}.
%\label{res2}\end{equation}
%\begin{equation}
%\begin{array}{c}
%\Gamma(\omega,T)=\displaystyle{\frac{1}{2^3\pi^2 \omega}\int d^3{\bf k}_1 w^2_{ph-ph} \omega_{\bf{ k}_1}[1+2N(\omega_{\bf{ k}_1})]\delta(2\omega_{\bf {k}_1}-\omega)}\,.\end {array}
%\label{res1}\end{equation}

\begin{figure}[h]
\resizebox{.5\textwidth}{!}{\includegraphics{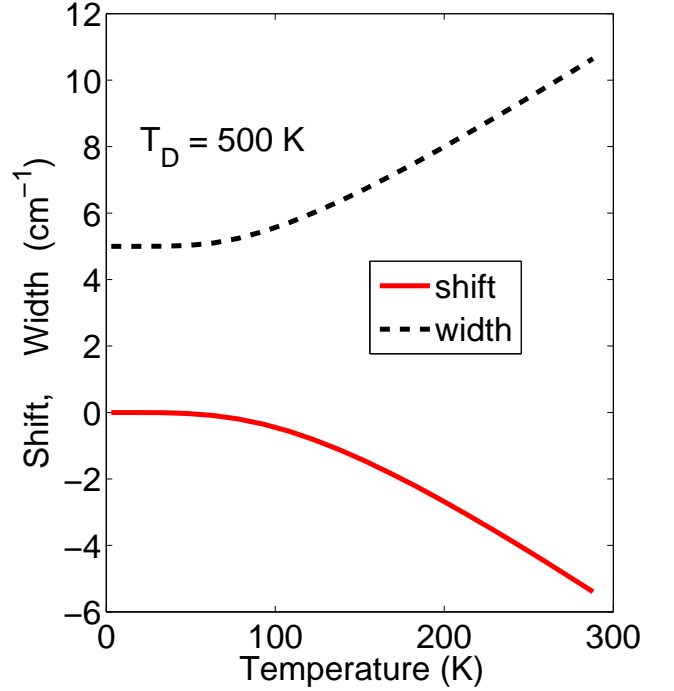}}%pw1.eps}}%xy7.eps
\caption{(Color online) Calculated width and shift of the Raman line $\omega=400 $ cm$^{-1}$ due to the anharmonic interaction of the third order as functions of temperature.}
\label{anh_wid1}
\end{figure} 

The Raman lineshift due to the phonon anharmonicity is given by the real part of Eq. (\ref{res31}), i. e. by the principal integral value 
\begin{equation}
\begin{array}{c}
\delta\omega_{ph-ph}(\omega,T)=\displaystyle{\frac{1}{2\pi^3 \omega}\dashint\frac{g^2_{ph-ph}\omega_{\bf{ k}_1}^3d\omega_{\bf{ k}_1} d\Omega}{s^5(4\omega_{\bf{ k}_1}^2-\omega^2)}[1+2N({\omega_{\bf{ k}_1}})]}\,.\end {array}
\label{res32}\end{equation}
%\begin{equation}
%\begin{array}{c}
%\delta\omega=\displaystyle{\frac{-1}{2(\pi \omega_D)^3}\dashint\frac{g^2_{ph-ph}\omega_{\bf{ k}_1}d\omega_{\bf{ k}_1} d\Omega}{\omega^2-4\omega_{\bf{ k}_1}^2}}[1+2N(\omega_{\bf{ k}_1})]\,,\end {array}
%\label{res32}\end{equation}
Here, the temperature dependent part appears because of the phonon distribution function $N({\omega_{\bf{ k}_1}})$.  At low temperatures,    $T\ll\omega/2 $,  we can omit $4\omega_{\bf{ k}_1}^2$, and the lineshift becomes
\begin{equation}
\begin{array}{c}
\delta\omega_{ph-ph}(\omega,T)=-\displaystyle{\frac{T^4}{\pi^3\omega^3}\int_0^{\omega_D/T}\frac{x^3dx}{e^x-1}\int\frac{g^2_{ph-ph}}{s^5}} d\Omega\,.\end {array}
\label{res41}\end{equation}
%\begin{equation}
%\begin{array}{c}
%\delta\omega(T)=-\displaystyle{\frac{T^2}{(\pi \omega_D)^3\omega^2}\int_0^{\infty}\frac{x^2dx}{e^x-1}\int g^2_{ph-ph} }d\Omega\,,\end {array}
%\label{res41}\end{equation}
 We  express  the lineshift in Eq. (\ref{res41}) in terms of the linewidth at zero temperature, Eq.
   (\ref{res2}),
\begin{equation}
\begin{array}{c}
\delta\omega_{ph-ph}(T)=-\displaystyle{\Gamma_{ph-ph}(\omega,0)\frac{2^6}{\pi}\left(\frac{T}{\omega}\right)^4\int_0^{\omega_D/T}\frac{x^3dx}{e^x-1}}\,.\end {array}
\label{res51}\end{equation}

The shift and width are shown in Fig. \ref{anh_wid1} for the Raman line 400 cm$^{-1}$ and the Debye temperature $T_D=500$ K   estimated for CoS$_2$. Men\'{e}ndez and Cardona \cite{MC} noted that the discrepancy between the calculated phonon linewidth  and experimental results may be of the order   of ten, "mainly from the poor description of the phonon dispersion curves". Therefore, Eq. (\ref{res51}) should be used in fitting with caution.  

Let us emphasize, that the calculated Raman linewidth due to three-phonon anharmonicity corresponds to the estimation
\begin{equation}\Gamma_{ph-ph}(\omega,0)\approx H_{ph-ph}^2/\omega\sim (m/M)^{1/4}\omega\sim 5\, \text{cm}^{-1}\,,\label{est1}\end{equation}
where  $m$ is the free electron mass, $M$ is the lattice cell mass, and the anharmonic  interaction has the order  \[H_{ph-ph}\approx \varepsilon_0 (u/a_0)^3\]
with $\varepsilon_0\sim \omega(M/m)^{1/2}\sim $ 3 eV  of the order of the atom energy. The ratio of the phonon displacement $u$ to the lattice constant $a_0
$ is of  $(m/M)^{1/4}$.
%\begin{equation}
%\Gamma(\omega,0)=\frac{\omega}{2^6\pi^2}\int \frac{g^2_{ph-ph}}{s^3}d\Omega\,,
%\label{res2}\end{equation}

%\begin{equation}
%\begin{array}{c}
%\delta\omega=\displaystyle{\frac{1}{4\pi^3}\dashint\frac{g^2_{ph-ph}\omega_{\bf{ %k}_1}^2d\omega_{\bf{ k}_1} d\Omega}{s^3(\omega^2-4\omega_{\bf{ k}_1}^2)}}[1+2N(\omega_{\bf{ %k}_1})]\,.\end {array}
%\label{res3}\end{equation}

\section{Effect of the phonon-magnon interaction on the optical phonon }

As seen from experimental data \cite{La}, besides  the phonon anharmonicity,  the ferromagnetic ordering in CoS$_2$ effects the Raman line form. The influence of antiferromagnetic ordering is considered in Ref. \cite{DPV}, however,   the line shift was only calculated.

\begin{figure}[b]
\resizebox{.5\textwidth}{!}{\includegraphics{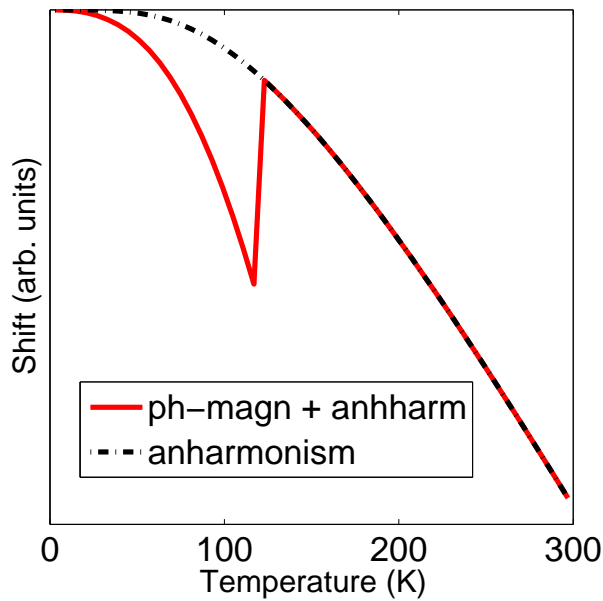}}%wid_mag.eps}}%pw1.eps}}%xy7.eps
\caption{(Color online) Temperature dependence of the Raman lineshift as a result of  phonon-magnon interactions.}%Температурная зависимость уширения  рамановской линии $\omega=400 $ см$^{-1}$ в результате фонон-магнонного взаимодействия.}
\label{ph-mag}
\end{figure}

 Here, we analyse the Raman shift and width due to the phonon-magnon interaction at  ferromagnetic ordering. The problem can be solved in the same manner as in the previous section considering the magnons instead of the acoustic phonons and taking into account that the magnons exist only below the Curie temperature. In the Matsubara technique, the magnon Green   function has the form  
\begin{equation}
G(\omega_n,{\bf k})=\frac{1}{-i\omega_n-\varepsilon_{\bf k}}\,,
\label{mf}\end{equation}
where the magnon dispersion law writes as  $ \varepsilon_{\bf k}=A(a_0k)^2$ for  temperatures in the range 1 K$\ll T< T_c$  with  a constant $A$ of the order of the Curie temperature. 
The summation should be carried out over frequencies $\omega_n=2\pi T n$ in the product of two magnon Green functions:
\begin{equation}
S_{ph-m}(\omega,{\bf k})=\sum_n\frac{-T}{(-i\omega_n-\varepsilon_ {\bf k1})[-i(\omega-\omega_n)-\varepsilon_{\bf k -k1}]}\,.
\label{ms1}\end{equation}
Similarly to the previous section, the summation and continuation to the real frequency give
\begin{equation}
S_{ph-m}(\omega,{\bf k})=\frac{1+N(\varepsilon_{\bf k1})+N(\varepsilon_{\bf k-k1})}{-\varepsilon_ {\bf k1}-\varepsilon_{\bf k -k1}+\omega+i\delta}\,.
\label{ms}\end{equation}

The Raman phonon self-energy is obtained integrating  Eq.
 (\ref{ms})  with  $\bf{k}=0$ over 
${\bf k_1}$ 
\begin{equation}
\begin{array}{c}
\Sigma_{ph-m}=\displaystyle{\frac{1}{8\pi^3}\int\frac{w_{ph-m}^2({\bf k_1}) d^3{\bf k_1}}{2\varepsilon_{\bf{ k}_1}-\omega-i\delta}}[1+2N(\varepsilon_{\bf{ k}_1})]\,,\end {array}
\label{mp1}\end{equation}
where $w_{ph-m}({\bf k_1}) =g_{ph-m}a_0k_1$ is  the phonon-magnon interaction  vertex with $g_{ph-m}\sim T_c/\sqrt{\omega Ma_0^2}\sim T_c(m/M)^{1/4}$. 

The imaginary part of Eq. (\ref{mp1}) vanishes for the phonon frequency $\omega\sim 600 $ K in CoS$_2$, because it  is higher than the maximal value of $ 2\varepsilon_{\bf{ k}_1}\sim2T_c\sim 250$ K. So, the real decay of the considered optical phonon into two magnons is forbidden.

However, if the phonon frequency is lower, the decay becomes possible producing the width
\begin{equation}
\Gamma_{ph-m}=\frac{\bar{g^2}_{ph-m}\omega^{3/2}}{4\pi (2A)^{5/2}}\left[1+\frac{2}{\exp(\omega/2T)-1}\right]\,.
\label{gmp1}\end{equation}

The  temperature dependence of the real part of Eq. (\ref{mp1})  giving  the Raman lineshift  can  be easily extracted as 
\begin{equation}
\delta\omega_{ph-m}(\omega,T)=\frac{-\bar{g^2}_{ph-m}
}{2\pi^2\omega}\left(\frac{T}{A}\right)^{5/2}\int_0^{\infty}\frac{x^{3/2}dx}{\exp{(x)}-1}\,,
\label{dmp}\end{equation}
where   we omit the magnon energy compared with
the phonon frequency, $\omega\gg 2\varepsilon_{\bf{ k}_1}\sim 2 T_c$.
 
In Fig. \ref{ph-mag}, the phonon shift, Eq. (\ref{dmp}), as a result of the phonon-magnon interaction is shown in   the solid line assuming that the magnon dispersion is given with the constant $A$ for all  temperatures below the Curie temperature. We see a sharp jump just at the Curie temperature. Such a behavior is not observed in experiments.
%Интеграл здесь равен 1.78.

Collecting the values of the vertexes, we can   estimate the result of the phonon-magnon interactions as
\begin{equation}
\begin{array}{c}\delta\omega_{ph-m}/\omega \sim (T_c/\omega)^2(m/M)^{1/2}\sim 0.3\, \text {cm}^{-1}\,,\end {array}
\label{est2}\end{equation}
i. e., it is ten times smaller than the effect of the anharmonicity, Eq. (\ref{est1}). The reason for this is obvious, this is the very low Curie temperature in CoS$_2$. Therefore, we  consider the electron-phonon mechanism of the Raman linewidth, which can exist in half-metals.

\section{Electron-phonon interactions at the ferromagnetic ordering }

\begin{figure}[]
\resizebox{.3\textwidth}{!}{\includegraphics{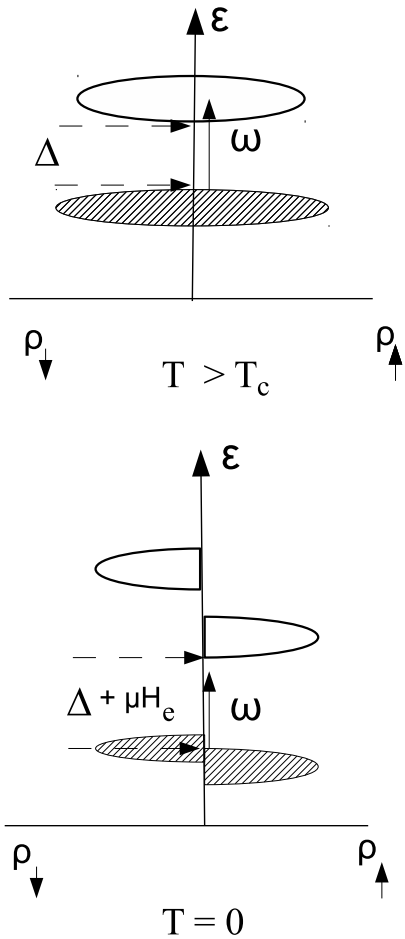}}%bandnn.eps}}
%xy7.eps
\caption{(Color online)  Proposed band scheme for two electron spin projections.}
\label{bands}
\end{figure}
We assume that the electron bands in CoS$_2$ have a form shown in Fig. \ref{bands}. The lower band is filled and the upper band is empty. At low temperatures, below the Curie temperature, the Raman frequency $\omega$ is less than the total gap, which consists of  the usual band gap $\Delta$ joined with the magnetic  splitting   $\mu H_e$ in the effective Weiss field  $ H_e$, where
$\mu$ is the total electron magneton  for the considered bands. In this case, the interband electron transitions are forbidden,  and the phonon width is determined only by the anharmonicity.  While the temperature increases, the magnetization,  determined in the mean field approximation as 
$$M=M_0\sqrt{1-(T/T_c)^2}\,,$$
becomes lower. Then the frequency $\omega$ can exceed the total gap $\Delta+\mu H_e$, and the interband transitions are possible. According to experimental data, this occurs in CoS$_2$ approximately at  $T=100$ K. The  proposed scheme  differs from the conventional electron-phonon interaction \cite{PLM,LPM,LM} only in the combination  of the interband electronic transition   with the ferromagnetic ordering. 

The phonon self-energy resulted from  the interband electron transition has the form
\begin{equation}
\Sigma_{el-ph}=g^2\sum_{n\neq m}\int\frac{d^3{\bf p}}{(2\pi )^3}\frac{f(\varepsilon_{{\bf p }m})-f(\varepsilon_{{\bf p }n})}{\varepsilon_{{\bf p }n}-\varepsilon_{{\bf p }m}-\omega-i\delta}\,, 
\label{d}\end{equation}
where the contribution of the electron transitions between the different $n, m$ bands is included, because we are interested in the phonon frequency $\omega\sim 400$  cm$^{-1}$, which is large  compared with the electron energy in half metals. The electron-phonon interaction vertex  $g$ is taken off the integrand, since it has no singularities in the electron energy interval considered. An estimation gives the vertex $g\sim \varepsilon_0(m/M)^{1/4}$, where $m$ and $M$ are the electron and ion masses, and  $\varepsilon_0\sim 3$ eV is the typical electron energy. The interval of values given in literature \cite{PLM} is
 $g$= 0.04$ \div $0.1 eV.

%\begin{figure}[]
%\resizebox{.5\textwidth}{!}%{\includegraphics{wi_anh_el-ph.eps}}%width4.eps}}%xy7.eps
%\caption{(Color online) Calculated temperature dependence of the Raman line width at the ferromagnetic transition with $T_c=120 K$; $g^2\tilde{m}^{3/2}\sqrt{\omega/2}/\pi=8$ cm$^{-1}$, $\mu H_e(T=0)=0.2\omega $, $\Delta=0.85 \omega$.}
%\label{width4}
%\end{figure}

The imaginary and real parts of the integral (\ref{d}) determine the variation of the phonon linewidth $\Gamma$ and  lineshift $\delta \omega$, correspondingly.  Results look  simply when the chemical potential  is situated in the gap and the temperature is much less than the gap, which value should be about the Raman phonon frequency 400 K. Then, the distribution function of the lower band is  $f(\varepsilon_{{\bf p }m})=1$ and for the upper band  $f(\varepsilon_{{\bf p }n})=0$.
 If the chemical potential belongs to any electron band, we have to add its value to the value of the gap. An additional simplification arises when the phonon frequency $\omega$ is close to the value of the gap $\Delta$. Then we can use the quadratic expansion  $$ \varepsilon_{{\bf p }n}-\varepsilon_{{\bf p }m}=\Delta+\mu H_e+p^2/2\tilde{m}$$ 
with the reduced mass $\tilde{m}$ of two bands.  

 \begin{figure}[h]
\resizebox{.5\textwidth}{!}{\includegraphics{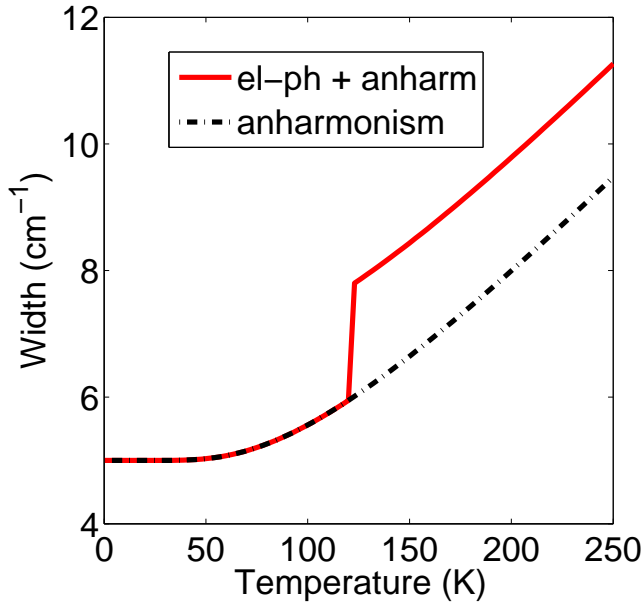}}%wid_anh_el1.eps}}%width4.eps}}%xy7.eps
\caption{(Color online) Calculated temperature dependence of linewidth for the Raman $\omega=400 $ cm$^{-1}$  line at the ferromagnetic ordering.}
\label{width4}
\end{figure}

Extracting the imaginary part and integrating  $\delta$-function in Eq. (\ref{d}), we find the width 
\begin{equation}
\Gamma_{el-ph}= \frac{g^2\tilde{m}}{2\pi}\sqrt{a}\,,\quad \text{for} \quad a>0\,,
\label{w}\end{equation}
where $a=2\tilde{m}(\omega-\Delta-\mu H_e)$. This contribution to the linewidth  vanishes, for  $a<0$. 

Taking the estimation of the vertex $g$ into account, we get the order of the width $\Gamma_{el-ph}/\omega\sim (m/M)^{1/4}/\pi\sim 3\times10^{-2}$, which corresponds with experimental data \cite{La}. The temperature dependence of the width is shown in Fig. \ref{width4} with the parameters % при этом взяты значения ширины за счет распада фонона при нуле температур $\Gamma_0=5.3$ см$^{-1}$, 
 $g^2\tilde{m}^{3/2}\sqrt{\omega/2}/\pi=8$ cm$^{-1}$, $\mu H_e(T=0)=0.5 \omega $, $\Delta = 0.95 \omega$, $\omega$ = 400 cm $^{-1}$. This value of $\Delta$ is in agreement with the gap $\sim 0.1$ eV calculated in Ref. \cite{ZCH}    at the $\Sigma$ direction in the Brillouin zone.

\begin{figure}[h]
\resizebox{.5\textwidth}{!}{\includegraphics{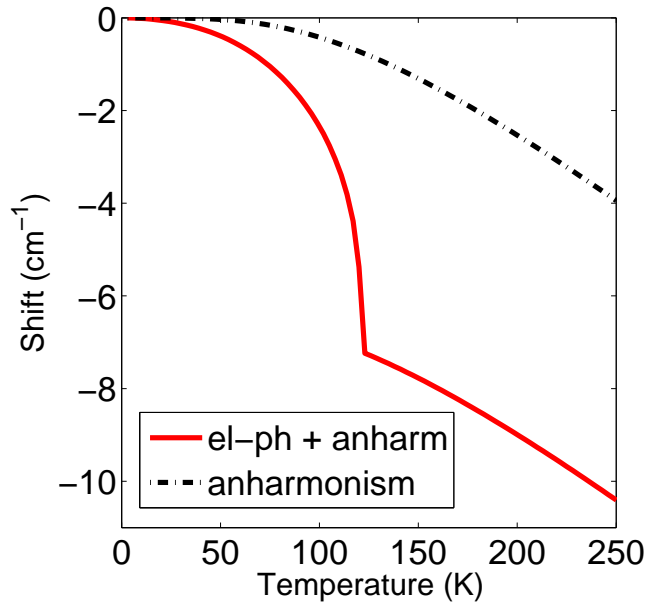}}%sh_anh_el-ph.eps}}%pos4.eps}}%xy7.eps
\caption{(Color online) Calculated shift of the Raman line $\omega=400 $ cm$^{-1}$ due to the electron-phonon interaction.}
\label{pos4}
\end{figure}

For the temperature dependence of the Raman lineshift, the integral (\ref{d}) gives
\begin{equation}
\begin{array}{c}
\delta\omega=-\displaystyle\frac{g^2\tilde{m}}{2\pi^2}\left[2\sqrt{b}\,\arctan{\frac{p_0}{\sqrt{b}}}\right.\\
+\left\{\begin{array}{c}\left.
\sqrt{a}\,\displaystyle\ln{\frac{p_0+\sqrt{a}}{|p_0-\sqrt{a}|}}\right]\,, \quad \text{for}\quad  a>0\,,\\
\left. 2\sqrt{|a|}\,\displaystyle\arctan{\frac{p_0}{\sqrt{|a|}}}\right]\,,\quad \text{for}\quad a<0\,,
\end{array}\right.
\end{array}
\label{w}\end{equation}
where $b=2\tilde{m}(\omega+\Delta+\mu H_e)$  and $p_0=20\sqrt{2\tilde{m}\omega}$ is the width  of the filled electron band in the momentum space. This temperature dependence of the lineshift is shown in Fig. \ref{pos4}.
%\begin{figure}[]
%\resizebox{.5\textwidth}{!}{\includegraphics{pos_anh_el1.eps}}%pos4.eps}}%xy7.eps
%\caption{(Color on line) Calculated shift $\omega=400 $ cm$^{-1}$ ;  $g^2\tilde{m}^{3/2}\sqrt{\omega/2}/\pi=10$ cm$^{-1}$, $\mu H_e(T=0)=0.2\omega $, $\Delta=0.85 \omega$.}
%\label{pos4}
%\end{figure}

There is an important distinction between Figs.  \ref{ph-mag} (phonon-magnon interactions) and \ref{pos4} (electron-phonon interactions).
The phonon-magnon interactions do not naturally influence  on the temperature dependence of the phonon width above the Curie temperature, whereas the electron-phonon interactions  are essential at higher temperatures because the interband electron transitions are possible at such the temperatures.
 The kinks in Figs. \ref{width4} and \ref{pos4} become smooth while the effect of the temperature on the electron distribution function is taken into consideration.
 
 \section{summary}
 The Klemens formula describes quantitatively the optical phonon width due to three-phonon anharmonic interactions. The corresponding lineshift matches with the width. However, the phonon-magnon interactions should be taken into account in order to interpret the effect of ferromagnetic ordering on the Raman line. In  such ferromagnets as CoS$_2$ with the low Curie temperature, these interactions are found to be too weak to describe quantitatively the experimental data. Therefore, we propose the mechanism of the electron-phonon interaction attended with the effect of the ferromagnetic ordering on the electron bands. The corresponding Raman line width and shift are calculated in  agreement with experiments.  
 \section{acknowledgments}
 The author thank S. Lapin and S. Stishov for information on their experiments prior the publication. 
This work was supported by the Russian Foundation for Basic
Research (grant No. 13-02-00244A) and the SIMTECH Program, New Centure of Superconductivity: Ideas, Materials and Technologies (grant No. 246937).

\end{document}